\documentclass[12pt,preprint]{aastex}

\usepackage{graphics,graphicx,epsfig}

\newcommand{\et}{et al.\ }

\newcommand{\ls}
{\mathrel{\hbox{\rlap{\hbox{\lower4pt\hbox{$\sim$}}}\hbox{$<$}}}}
\newcommand{\gs}
{\mathrel{\hbox{\rlap{\hbox{\lower4pt\hbox{$\sim$}}}\hbox{$>$}}}}

\newcommand{\xmm}{{\it XMM-Newton}}
\newcommand{\asca}{{\it ASCA}}
\newcommand{\sax}{{\it BeppoSAX}}
\newcommand{\pks}{PKS~2155--304}
\newcommand{\mkn}{Mkn~421}
\newcommand{\mos}{EPIC--MOS}
\newcommand{\pn}{EPIC--pn}

\begin{document}

\title{ Complex X-ray spectral variability in \mkn\ observed with \xmm}

\shorttitle{ \xmm\ Monitoring of \mkn}
\shortauthors{ Sembay \et }

\author{S. Sembay\altaffilmark{1} ,
        R. Edelson\altaffilmark{1,2},
        A. Markowitz\altaffilmark{2},
        R.\ G.\ Griffiths\altaffilmark{1}, and
        M.\ J.\ L.\ Turner\altaffilmark{1} }

\email{sse@star.le.ac.uk}

\altaffiltext{1}{X-ray Astronomy Group; Leicester University; Leicester
LE1 7RH; United Kingdom}

\altaffiltext{2}{Astronomy Department; University of California; Los
Angeles, CA 90095-1562; USA}

\begin{abstract}

The bright blazar \mkn\ has been observed four times for uninterrupted
durations of $ \sim 9 - 13 $~hr during the performance verification and
calibration phases of the \xmm\ mission.
The source was strongly variable in all epochs, with variability
amplitudes that generally increased to higher energy bands.
Although the detailed relationship between soft (0.1--0.75~keV) and hard
(2--10~keV) band differed from one epoch to the next, in no case was there
any evidence for a measurable interband lag, with robust upper limits of $
| \tau | < 0.08 $~hr in the best-correlated light curves.
This is in conflict with previous claims of both hard and soft lags of
$\sim$1~hr in this and other blazars.
However, previous observations suffered a repeated 1.6~hr feature induced
by the low-Earth orbital period, a feature that is not present in the
uninterrupted \xmm\ data.
The new upper limit on $|\tau|$ leads to a lower limit on the magnetic
field strength and Doppler factor of $ B \delta^{1/3} \gs 4.7 $~G, mildly
out of line with the predictions from a variety of homogeneous synchrotron
self-Compton emission models in the literature of $ B \delta^{1/3} = 0.2 -
0.8 $ ~G.
Time-dependent spectral fitting was performed on all epochs, and no
detectable spectral hysteresis was seen.
We note however that the source exhibited significantly different spectral
evolutionary behavior from one epoch to the next, with the strongest
correlations in the first and last and an actual divergance between soft
and hard X-ray bands in the third.
This indicates that the range of spectral variability behavior in \mkn\ is
not fully described in these short snippets; significantly longer
uninterrupted light curves are required, and can be obtained with \xmm.
\end{abstract}


\section{ Introduction }

\mkn\ has a long history of observations at X--ray and other wavelengths.
It
is nearby ($\rm z = 0.031$) and is generally the brightest BL Lacertae
object in the X--ray sky. BL Lac objects are a sub--class of blazars;
sources whose broad--band continuum emission are dominated by relatively
featureless nonthermal radiation and which show rapid spectral variability
with the characteristic timescale of variations typically decreasing with
increasing energy.

The spectral energy distribution (SEDs) of blazars have two distinct peaks
in $\nu F_{\nu}$ space; one in the IR/optical or UV/X--ray regime and one
at
higher energies in the X--ray/Gamma--ray regime. The lower energy
component is probably incoherent synchrotron radiation and the high energy
component
Compton up--scattering of lower energy photons. It seems likely that
the
synchrotron photons provide the seeds for the Compton emission (the
synchrotron self--Compton process), but this has not been unambiguously
demonstrated. BL Lacs are often classified as
LBLs (low--energy peaked BL Lacs) if the synchrotron component peaks in
the
IR/optical regime or HBLs if it peaks in the UV/X--ray, although the
classes
may be the extrema of a single population (e.g. Fossati \et 1998). \mkn\
is an HBL.

The rapid flux variability in blazars would require the luminosity to be
highly super--Eddington if the emission was isotropic and it is generally
assumed that the emission region moves at relativistic speeds towards the
observer; an interpretation supported by observations of superliminal
motion of radio knots in the jet--like structure of some radio galaxies.

Although the relativistic jet model provides a plausible
explanation for the continuum properties of blazars, detailed physical
constraints on the structure of the jets remain an elusive goal;
{\it snap--shot} multi--wavelength spectra can be reproduced by a variety
of
jet models with different underlying assumptions. Studying the variability
behaviour of both the spectrum and flux is therefore the most
profitable way to constrain the various jet models. The relationship of
the
variations within different energy bands can, in principle, help to
constrain the geometry of the emission region on the macroscopic
scale and constrain the physical radiation and loss processes on the
microscopic scale. For instance, X--ray variations in the HBL, \pks, have
been seen to lead those in the optical/UV (e.g., Edelson \et 1995;
Urry \et 1997), supporting the view that the X-rays arise from the
synchrotron
component and not from Compton scattering from these lower--energy bands.

The X--ray spectral slope is especially sensitive to electron cooling
timescales within the various models and observations of interband lags 
within this regime have often been interpreted in this context. 
Chiappetti \et (1999) found that soft X--ray (0.1--1.5 keV) variations in 
\pks\ led hard
X--ray (3.5--10~keV) variations by $\sim$0.3-4~hr and suggested it could
occur through radiative cooling if the electron population has a sharp
break or low--energy cut--off. Historically, however, the source has not
exhibited a consistent behaviour in this respect.
X--ray flares have also been observed with the hard band leading the soft
(Sembay \et 1993).

\mkn\ has also been the subject of a number of monitoring campaigns
and shows a similar variety of behaviour between observations. Takahashi
\et (1996) found that the hard X--ray band led the soft, whereas
Fossati \et (2000a) found the soft band led the hard. Takahashi \et (2000)
reported both types of behaviour seen within a single very long 
observation ($\sim 11$~days) of the source by \sax\ and \asca.

Interpretation of interband lags can, however, be complicated by the fact 
that these observations have been made with low--Earth orbit satellites, 
resulting in periodic interruptions in the light curves every $\sim 1.6$ hr 
due to Earth--occultation. Edelson \et (2001; hereafter E01) analysed 
uninterrupted \xmm\ data on \pks\ and found no significant interband lag 
down to a reliable limit of $\sim 0.3$ hr, attributing the previous claims 
of lags on short time scales to a spurious signal induced into the CCF at the 
orbital period of the satellites. Takahashi \et (2000) also 
noted the possibility of such an effect in their 
paper on \mkn. One advantage \xmm\ has over previous X--ray
missions for monitoring purposes is that it has a highly eccentric orbit
with
a 48 hour period, thus allowing much longer continuous observation
periods.
Nominally the only data gaps occur during passage through the radiation
belts
at perigee which takes around 7--8 hrs and, in the first year or so of the
mission, a $\sim 1$ hour gap at apogee due to incomplete ground station
coverage (Jansen \et 2001).

\xmm\ has observed both \pks\ and \mkn\ on several occasions since launch.
We have reported on the \xmm\ observation of \pks\ in E01.
In this paper we now present the results of several observations of \mkn.
The observations and data reduction are discussed in the next section. The
temporal analysis of these data are reported and compared with other data
in \S~3. The spectral analysis is discussed in \S~4 and a concluding 
discussion is given in \S~5.

\section{ Observations and Data Analysis }

\subsection{ Observations }

The \xmm\ observations of \mkn\ were performed as part of the 
performance--verification and calibration programme. The source was observed
on--axis on four occasions at the epochs listed in Table~1. \xmm\ has
three
co--aligned X--ray telescopes with focal plane cameras of two types; two
\mos\ imaging CCD cameras and one \pn\ imaging CCD camera (Turner \et
2001;
Str\"{u}der \et 2001). Within the telescopes containing the MOS cameras,
gratings divert approximately half the incident flux to two Reflection
Grating Spectrometer (RGS) instruments (den Herder \et 2001).

In this paper our analysis concentrates on data taken with the
high--throughput EPIC CCD cameras. The EPIC cameras have superior
sensitivity
to short timescale flux variations and a broader energy response than the
RGS
instruments. During orbital revolutions 165, 171 and 259 (henceforth, O165, 
O171 and O259) both EPIC cameras were in their
respective Small Window (SW) imaging modes throughout the observation.
In orbit~84 (O84), however, the first part of the observation was 
performed in
timing mode for the pn and MOS1 cameras, and SW mode for MOS2, followed by
SW mode for the pn and MOS1 and timing mode for MOS2. There is, however,
only a small overlap of about 5 ksec between the pn and MOS imaging data.
Due to a radiation alert the MOS cameras were switched off shortly after
the beginning of the pn SW mode observation (further details of the 
O84 observation can be found in Brinkmann \et (2001) which gives the 
first analysis of the data from this observation). A variety of blocking 
filters were employed throughout the observations and these are listed in 
Table~1.

The data analysis software and calibration were generally in a more mature
state for imaging mode than timing mode when the analysis was performed,
so
this paper deals with the imaging data only. 

The imaging observations were performed in each instrument's SW mode
because
they have a faster CCD frame readout time than their respective
Full--Frame
modes. This mode reduces the effects of pulse pile--up which causes
spectral distortion in bright sources. Pulse pile--up occurs when two (or more)
X--rays interact within a given frame close enough such that their
deposited
charge distributions overlap. This can lead to a loss of flux if the
resultant charge distribution does not have a valid pattern within the
instruments X--ray pattern library, or, if it still has a valid pattern, a
new event whose energy is the sum of the contributing events. Hence,
pile--up
leads to an artifical hardening of the spectrum and can produce spurious
correlations between soft and hard--band variations. The simplest way to 
correct for pile--up is to exclude the core of the point--spread function 
where pile--up is greatest. This is explained in more detail in the 
following Section.

The pn SW mode has a $63 \times 64$ pixel window (equivalent to $\sim 4.3
\times 4.4$ arcminutes on the sky) with a frame readout time of 5.7~ms.
The
common telescope boresight places the source off--centre within the
window,
typically less than an arcminute from the CCD boundary. For the MOS
cameras
the window is $100 \times 100$ pixels ($\sim 1.8 \times 1.8$ arcminutes),
more centered with respect to the boresight, and the mode has a frame
readout time of 0.3~s. 

\placetable{table1}

\subsection{ Data Reduction }

A standard reduction of the raw pn and MOS event lists was performed using the
Science Analysis System (SAS) Version 5.2. The SAS removes events from the 
location of defective pixels, corrects for charge transfer losses across the 
CCDs and applies a gain calibration to convert the recorded charge in ADC units
to a pulse--invariant energy scale. Standard analysis packages are then
used to create images, light--curves and spectra from the calibrated event
lists. 

Pulse pile--up in CCD detectors is nominally a purely statistical effect
which can in principle be corrected for by modifying the instrument
response
function appropriately. However, without having a suitable tool available
within the SAS, we have used the more direct approach of excluding events
from the core of the point--spread function where, naturally, pile--up is
the
most extreme.  Although this leads to a loss of flux, the high throughput
of
XMM means that we are still left with spectra and light curves with high
statistical significance. 

The size of the core to be excluded was determined by examining the
distribution of mono--pixel to multi--pixel events as a function of
energy.
As the radius of the excluded region is increased, the distribution tends
towards that consistent with the theoretical distribution for the
non piled--up case. By this method we excluded core radii of diameter
8'', 6'', 10'' and 10'' respectively for the pn O84, O165, O171
and O259 observations
and 8'', 12'' and 12'' for the MOS O165, O171 and O259 observations. The 
maximum  radius of each annulus was 36.75'' during O84 increasing to 44'' 
for the remaining observations; the difference being due to a change in the 
relative boresight with respect to the pn window.

The point-spread function of each telescope is energy dependant so a 
correction needs to be applied to the effective area of the response file 
when fitting spectra extracted from an annulus with a gven inner and outer
radius. To do this we used a description of the psf developed by the
EPIC instrument teams using in-flight data from several on-axis point sources
(Ghizzardi \& Molendi 2002). The robustness of this procedure and other 
calibration related issues are discussed in more detail in Section 4.

In table~1 we list the mean observed $0.1-10.0$ keV source count rates 
(obtained {\it after} excluding the psf cores) during each observation. The 
source was most luminous during O171, but only $\sim 1/3$ as bright twelve 
days earlier during O165. O84 and O259 are at intermediate levels. 
The higher observed count rate in the soft band during O259 is primarily a
consequence of the higher throughput of the Thin blocking filter compared 
with the Medium and Thick filters used earlier.

\section{ Timing Analysis }

In this section, we compare gross variability in different energy bands
for each of the observing epochs. Much of this analysis parallels 
that on \pks\ by E01. Although most of the analyses will be restricted to 
higher signal-to-noise pn data, the relevant MOS data are also included for 
consistency.

The light curves were constructed as follows: first, data were divided
into soft (S; 0.1--0.75~keV), medium (M; 0.9--1.7~keV) and hard (H;
2--10~keV) bands. Then, the data were initially extracted at 1~sec intervals.
They were then rebinned to 300~sec resolution and standard methods were
used to measure the mean and standard errors from the spread of data in
each bin. Due to data loss, there were two small gaps in the O165 light
curves. The four points during 1.75--2.00 hrs and three points during
5.08--5.25 hrs after the beginning of the integration were instead
determined by interpolation from the surrounding points.

In all bands, the source count rate exceeded the background by at least a
factor of $\sim 50-100$, so no background light curves were extracted.
The resulting light curves are shown in Figure~1.

\subsection{Variability Amplitudes}

Variability amplitudes were measured for each light curve using the
fractional variability amplitude, defined as
\begin{equation}
f_{v} = \frac{\sqrt{S^{2} - \langle \sigma^{2}_{err} \rangle}}{\langle X \rangle} ,
\end{equation}
where $S^2$ is the total measured variance, $\langle \sigma^{2}_{err} \rangle$ is the mean squared errors, and $\langle X
\rangle$ is the mean count rate (see E01 for details). The fractional 
variability amplitude is also simply the square-root of
the ``excess variance" parameter used e.g., in Nandra \et (1999); $ f_{v} =
\sqrt{{\sigma^{2}}_{xs}} $. These quantities are computed and shown in 
Table~2. Examination of this and Figure~1 indicate that stronger variations are
clearly seen in the harder bands.
This increasing variability with energy band is common in blazars (e.g.,
Ulrich, Maraschi \& Urry 1997) and is a consequence of the fact 
that high energy electrons have shorter cooling timescales than low energy
ones. Finally, different variability levels are seen in each of the epochs with
O84 exhibiting the greatest overall degree of flux variations.

\subsection{Comparison of the Hard and Soft Light Curves}

Figure~2 shows a variety of quantities derived directly from the light
curves in Figure~1.
The top panel gives the geometric mean flux of the three bands,
$ F_g = \sqrt[3]{ F_H \times F_M \times F_S } $.
The second panel is the hardness ratio, $ HR = F_H / F_S $.
Note that the mean and hardness ratio light curves generally track each
other very well, confirming that the source generally gets harder as it gets
brighter, with the exception of the end of O165 where we observe a 
softening of the source as it continues to brighten.

Below that is an ``overplot" diagram for the hard and soft bands.
In these the light curves are subjected to the first two steps of
cross-correlation function (CCF) analysis: at each point in the light
curve, the mean count rate is subtracted and then the data are divided by
the standard deviation.
The resulting plot, with zero mean and unit variance for each light curve,
provides a both a simple way to compare variations between bands as well
as a convenient ``sanity check" of the final CCFs by showing the
intermediate step. Finally, the bottom panel is the difference between the 
two scaled light curves. 

The temporal relation between hard and soft band variations shows a range
of behaviors: O84: almost perfect agreement to within the errors; O165 and
259: good general agreement but significant divergances on short
($\gs$hour) time scales; O171: poor overall agreement and even a
divergence between soft and hard bands.
As seen below, the light curves that track the best (O84 and O259) are
particularly useful because they allow the finest measurement/limits on
the interband lag.

\subsection{Cross-Correlation Analysis}

Cross-correlation functions were measured using the discrete correlation
function (DCF; Edelson \& Krolik 1989) and the interpolated correlation
function (ICF; White \& Peterson 1994).
These are shown in Figure~3, where a positive $\tau$ indicates that the 
softer band leads the harder band. They are also tabulated in Table~3.
The errors, measured using the bootstrap method of Peterson \et (1998),
indicate a 2$\sigma$ upper limit of $ \tau \le 0.08 $~hr.

The best-tracking epoch (O84) also showed the highest correlation
coefficients ($ r \ge 0.99 $) and the strongest limits on the interband
lags ($ \bar{\tau} = -0.01 \pm 0.03 $).
Such similar limits are seen in the next two strongest correlations (O165
S/M and O259 M/H, with $ r = 0.96 $) and for the eight highest
correlations (out of 12, those with $ r \ge 0.89, \bar{\tau} = -0.01
\pm 0.04 $).
Recall that the bin size is 0.08~hr (300~sec); these indicate very secure
($2-3\sigma$) limits of $ | \tau | < 0.08 $~hr).

However, it is also clear especially in O171 that the hard and soft light
curves are not always tracking each other with high precision.
In this case, the hard and soft actually diverge, and the peak correlation
function is less than zero.
This reveals a further complexity in the spectral variability that cannot
be addressed with these short snippets of data, and indicates the need for
much longer uninterrupted observations to fully describe the observational
picture.

\subsection{Comparison with Previous Results}

Previous observations of \mkn\ have claimed significant hard lags, in
which the hard band lagged behind the soft by $\sim$0.6~hr (Fossati \et
2000a), as well as soft lags, in which the soft band lagged behind the
hard by $\sim$1~hr. By comparison, this paper finds that only during one of 
the four epochs are the data sufficiently well correlated to measure a 
meaningful CCF, and in that case, there is no measurable interband lag down 
to much tighter 2$\sigma$ limits: $ |\tau| \le 0.08 $~hr (1 bin width).

In order to understand this discrepancy, it is important to keep in mind
that the previous claims of interband lags all involved observations made
with low-Earth orbit satellites (\sax\ and \asca).
As pointed out by E01, CCFs measured from such data are
corrupted on time scales less than the orbital period (1.6~hr) and thus
can yield spurious apparent lags.
This can be seen in Figure~11 of Fossati \et (2000a): there is a periodic
feature running through the CCF with a period of 1.6~hr, causing
excursions larger than any claimed local peaks.
Indeed, the highest point in that CCF is at zero lag, and the next two
highest points are immediately beside it.
On the other hand, \xmm\ produces uninterrupted light curves that do not
suffer from this systematic errors, so the CCFs are intrinsically more
reliable.
Thus, we conclude that there is no reliable evidence for interband lags,
hard or soft, in \mkn.

\section{ Spectral Analysis }

The instrinsic X--ray spectra of BL Lacs are basically smooth and
featureless. Over an energy band of roughly a decade the continuum can be 
well represented by a power law model whereas over a wider band the 
continuum exhibits a degree of curvature requiring fitting by
broken--power law or continuously curved (CC) spectral models 
(e.g. Inoue \& Takahara 1996; Fossati \et 2000b). In this paper we have 
confined our analysis to the high energy 2.0 to 10 keV spectral band. There 
are unresolved cross--calibration issues between the EPIC 
instruments in the soft band which lead to somewhat contradictory results 
regarding the slope of the low--energy continuum. A full analysis of the 
entire spectral range will be left to a later paper. In the chosen band, 
the derived spectra can be fit a single power law with the column density 
fixed at the Galactic line--of--sight value of 
$\rm 1.6 \times 10^{20} \ cm^{2}$ (Lockman \& Savage 1995) which is in 
agreement with previous spectral analyses of the broad--band X--ray spectrum 
(e.g. Fossati \et 2000b).

Our spectral analysis on Mrk~421 proceeded as follows. Having removed the
piled--up section of the point spread function as described earlier, we
subdvided each observation into 5~ksec segments. This allowed us to
extract spectra with good statistical significance and still track spectral
variations throughout each observation. Between six and nine spectra were
extracted per observation and per camera. For completeness, background 
spectra were extracted, but the background rate was found to be negligible 
with respect to the source and hence the background spectra were ignored. The 
pn data were analysed separately to the MOS, but the data from both MOS cameras
were combined in the spectral fitting. All spectral fitting was done with 
Xspec Version 11.00. 

Table~4 lists the derived spectral parameters from each fit. There is an 
excellent agreement between the pn and MOS parameters in general. The pn 
photon indeces for example are typically within 
$\Delta \alpha \sim 0.05$ of the MOS. The overal range in 2-10~keV source flux
is similar to that observed in the \sax\ campaigns of 1997 and 1998 
(Fossati \et 2000b).

In Figure~4 we have plotted the spectral index from all observations, 
against the 2 to 10 keV (deabsorbed) flux. For clarity, the epic-pn parameters
only are plotted. From this plot we can see that although the variation of 
slope with flux follows in general the classical trend of synchrotron 
emission, namely, that steeper slopes correspond with lower flux states, 
there is no strong correlation between these parameters globally. As expected 
from the previous discussion of the light curves, the variability behaviour 
is quite different between observations. In O84 the slope is stable but the 
flux varies. In O165, when the source was weakest and the spectrum steepest, 
neither parameter varied strongly. O171 and O259 show a similar behaviour in 
that a fairly linear correlation between slope and flux is evident.

\section{ Discussion }

The temporal analysis found that in O84, the data were highly
correlated with no evidence for an interband lag down to time scales of 
$|\tau| \le 0.08 $~hr, much tighter limits than previous claims of both hard
and soft lags. Following on from similar results on \pks\ by E01 this
provides further strong evidence that the detection of lags on short 
time scales by low--Earth orbit satellites are probably spurious artefacts 
caused by Earth occultation.

Measurements of a lag could potentially give the cooling time scale and 
therefore a limit on the magnetic field strength assuming that the lag is 
due to the electron population softening due to radiative losses. From 
Chiappetti \et (1999),

\begin{equation}
B \delta^{1/3} = 300 \left( \frac{1+z}{\nu_{1}} \right)^{1/3} 
\left( \frac{1 - ( \nu_{1} / \nu_{0} )^{1/2} }{\tau} \right)^{2/3} G,
\end{equation}

where $B$ is the magnetic field strength, $\delta$ is the Doppler
factor, $z$ is the redshift, and $\nu_{1}$ and $\nu_{0}$ are the frequencies
(in units of $10^{17}$~Hz) at which a lag, $\tau$, is measured. 
For the Orbit 84 measurement, we have an upper limit on the lag so we can 
derive a lower limit on the value of $B \delta^{1/3}$. The most stringent 
limit is given by the hard and soft light curves with $\nu_{0} \sim 14.5$ and
$\nu_{1} \sim 1.0$, $z = 0.031$ and 
$\tau \stackrel{<}{\scriptstyle \sim} 300$s giving 
$B \delta^{1/3} \stackrel{>}{\scriptstyle \sim} 4.7$~G.

This lower limit is rather higher (typically by a factor of $\sim 5-10$) than
obtained from some estimates of $B$ and $\delta$ in the literature derived 
from fitting a variety of homogeneous SSC models to measurements of the 
broad--band SED of Mkn~421 observed at different epochs. For example, 
Ghisellini \et (1998) obtain $B = 0.09$~G and $\delta = 12$ 
giving $B \delta^{1/3} = 0.2$~G for a pure SSC model. If the seed photons of 
the Compton emission arise from an external source (e.g. an accretion disk or 
the broad--line region) then the model allows for a somewhat higher magnetic 
field strength. In this case Ghisellini \et (1998) obtain $B = 0.22$~G and 
$\delta = 11$ giving $B \delta^{1/3} = 0.49$~G. In comparison, Tavecchio, 
Maraschi and Ghisellini (1998), obtained $B = 0.25$~G and $\delta = 25$ giving 
$B \delta^{1/3} = 0.7$~G, also with a homogeneous SSC model, and 
Krawczynski \et (2002) argue for $B = 0.22$~G and $\delta \sim 50$ which 
gives $B \delta^{1/3} = 0.8$~G. 

Despite the differences in the underlying assumptions behind the specific
homogeneous models employed a rather narrow range of values of 
$B \delta^{1/3}$ are predicted. It is possible that source variability
is the cause of the discrepancy with, perhaps, the magnetic field strength
during O84 being unusually high compared with earlier epochs. Unfortunately 
we have no way to directly test this because we have no simultaneous 
observations in the gamma-ray regime with the O84 observation. The 
Krawczynski \et (2002) paper does deal with simultaneous X-ray and TeV 
gamma-ray observations (obtained with {\it RXTE}, the {\it Rossi X-ray Timing 
Explorer} and HEGRA, the High Energy Gamma Ray Astronomy experiment) taken in 
February and 3-8th May 2000, which is close to, but not contemporaneous with 
the 25th May 2000 O84 observation. Appealing to source variability, however,
seems contrived given the fact that observations from different epochs
are returning a similar range of magnetic field strengths. The previous 
claims of lags on longer, $\sim 1$~hr, timescales lead to 
values of the $B$ and $\delta$ combination that are consistent with these 
models and this has been used as an argument in their favour on the grounds of 
self-consistancy. Our much tighter upper limit on the lag may, however, 
indicate a possible incompatibility with the homogeneous SSC model although 
the results are not conclusive. 

It interesting that even in these short \xmm\ observations the spectral 
variability behaviour of \mkn\ is quite complex. Of special interest is the 
lack of any obvious spectral hysteresis in the variation of intensity with 
slope as has been commonly observed in a number of blazars, including \mkn\ 
(e.g. Sembay \et 1993, Takahashi \et 1996). When observed it 
most commonly follows a clockwise loop in the sense of Figure~4. This
behaviour naturally follows whenever the slope is controlled by cooling
processes which act faster at higher energies, as is the case when 
synchrotron cooling dominates. If the system is in a regime where the cooling
and acceleration times are equal, however, the reverse behavour is expected 
and the slope will track the intensity in an anticlockwise loop (Kirk, 
Rieger \& Mastichiadis 1998). 

There are a number of possible reasons why our observations show no clean
examples of hysteresis. First, it may simply be that the observations are
too short relative to the characteristic flaring timescale and therefore
we are observing only partial segments of a given loop. This may explain
the difference in the 2-10 keV spectral behaviour between O84 and O171
which both show a decay in flux. In the latter case the observation may be 
starting from near the peak intensity where a rapid change in slope would be 
expected during the initial decay. In the former we may be well below the 
peak intensity in a segment where we expect only a minor change in slope 
(see Figure~3 from Kirk, Rieger \& Mastichiadis 1998). Second, there may be 
contributions from multiple flares at different phases of their cycle; the 
lack of a correlation between the soft and hard bands in O171 appears to show 
an independent soft flaring component. Finally, the system overall may 
physically be in a transition zone between characteristically clockwise and 
anticlockwise spectral behaviour.

Attempts have been made to monitor \mkn\ and other BL Lacs with long 
observations by \sax\ and \asca\, however, these 
are limited on what they could achieve on analysing short time scales by 
being interrupted by Earth occultation. \xmm\ can provide the uninterrupted 
light curves on the short time scales required to unambiguously test for the 
presence of X--ray lags, but a source such as \mkn\ needs to be monitored for 
longer than the observations presented in this paper (which were designed 
primarily for instrument calibration purposes) in order to track the source 
behaviour over a wider range of time scales.

\acknowledgments

The authors thank the \xmm\ team for the huge amount of effort required to
build and operate this complex and powerful instrument. Sembay, Griffiths and 
Turner acknowledge financial support from the UK Particle Physics and 
Astronomy Research Council. Edelson and Markowitz are supported by NASA 
grants NAG~5-7317 and NAG~5-9023.

\begin{deluxetable}{lccccc}
\tablewidth{5.7in}
\tablenum{1}
\tablecaption{ Observation Summary \label{tab1}}
\small
\tablehead{
\colhead{ Orbit } & \colhead{ Inst. } & \colhead{ Observing Time } &
\colhead{ Mode } & \colhead{ Filter } & \colhead{ 0.1--10 keV} \\
\colhead{ } & \colhead{ } & \colhead{ 2000 UT } &
\colhead{ } & \colhead{ } &
\colhead{ $\langle$ $\rm cts \ s^{-1}$ } $\rangle$}
\startdata
84  &  pn   & 05 25 10:35:19 -- 05 25 19:36:58 & SW & Medium  & 88.65 \\
165 &  pn   & 11 01 23:48:12 -- 11 02 10:04:14 & SW & Thick   & 51.07 \\
171 &  pn   & 11 13 22:00:50 -- 11 13 10:57:29 & SW & Thick   & 125.23 \\
259 &  pn   & 05 08 09:31:39 -- 05 08 19:58:33 & SW & Thin    & 130.82 \\
165 &  MOS1 & 11 01 23:47:12 -- 11 02 10:09:02 & SW & Medium  & 18.87 \\
165 &  MOS2 & 11 01 23:47:12 -- 11 02 10:09:00 & SW & Medium  & 18.62 \\
171 &  MOS1 & 11 13 22:00:52 -- 11 14 11:11:37 & SW & Medium  & 38.52 \\
171 &  MOS2 & 11-13 22:00:53 -- 11 14 11:11:38 & SW & Medium  & 37.93 \\
259 &  MOS1 & 05 08 09:16:05 -- 05 08 19:56:13 & SW & Thin    & 31.25 \\
259 &  MOS2 & 05 08 09:16:04 -- 05 08 19:56:13 & SW & Thin    & 31.64 \\
\enddata
\end{deluxetable}

\begin{deluxetable}{llccrrr}
\tablewidth{5.3in}
\tablenum{2}
\tablecaption{ Variability Parameters \label{tab2}}
\small
\tablehead{
\colhead{ Orbit } & \colhead{ Band } & \colhead{ Npts } &
\colhead{ $\langle X \rangle$ } & \colhead{ $\sigma^{2}_{xs}$ } &
\colhead{ $\rm F_{var}^{1}$ } & \colhead{ S/N } \\
\colhead{ } & \colhead{ } & \colhead{ } &
\colhead{ $\rm cts \ s^{-1}$ } & \colhead{ $\times 10^{-3}$ } &
\colhead{ } & \colhead{ } }
\startdata
 84 & pn-S & 109 & 41.82 & 4.95 $\pm$ 0.34 &  7.03 $\pm$ 0.49 \%  & 106.10\\
 84 & pn-M & 109 & 24.44 &12.78 $\pm$ 0.88 & 11.31 $\pm$ 0.78 \%  & 81.41\\
 84 & pn-H & 109 & 10.62 &23.71 $\pm$ 1.63 & 15.40 $\pm$ 1.06 \%  & 53.59\\
\tableline
165 & pn-S & 123 & 26.97 & 5.96 $\pm$ 0.39 &  7.72 $\pm$ 0.52 \%  & 87.20\\
165 & pn-M & 123 & 13.11 & 7.16 $\pm$ 0.47 &  8.46 $\pm$ 0.56 \%  & 60.44\\
165 & pn-H & 123 &  4.55 & 7.71 $\pm$ 0.54 &  8.78 $\pm$ 0.62 \%  & 36.01\\
\tableline
171 & pn-S & 156 & 58.08 & 0.62 $\pm$ 0.04 &  2.49 $\pm$ 0.16 \%  & 114.13\\
171 & pn-M & 156 & 35.21 & 1.52 $\pm$ 0.09 &  3.90 $\pm$ 0.24 \%  & 88.59\\
171 & pn-H & 156 & 14.92 &14.23 $\pm$ 0.82 & 11.93 $\pm$ 0.69 \%  & 58.82\\
\tableline
259 & pn-S & 126 & 75.74 & 1.31 $\pm$ 0.09 &  3.62 $\pm$ 0.24 \%  & 147.81\\  
259 & pn-M & 126 & 28.83 & 4.22 $\pm$ 0.27 &  6.50 $\pm$ 0.42 \%  & 90.91\\ 
259 & pn-H & 126 & 11.60 &14.44 $\pm$ 0.93 & 12.02 $\pm$ 0.77 \%  & 58.36\\
\tableline
\tableline
165 & MOS-S & 125& 18.23 & 5.37 $\pm$ 0.36 &  7.33 $\pm$ 0.49 \%  & 64.62 \\
165 & MOS-M & 125& 10.89 & 7.44 $\pm$ 0.49 &  8.62 $\pm$ 0.57 \%  & 53.22 \\
165 & MOS-H & 125&  4.47 & 4.22 $\pm$ 0.31 &  6.50 $\pm$ 0.48 \%  & 39.22 \\
\tableline
171 & MOS-S & 157& 31.23 & 1.02 $\pm$ 0.07 &  3.20 $\pm$ 0.21 \%  & 84.39\\
171 & MOS-M & 157& 24.51 & 1.10 $\pm$ 0.07 &  3.32 $\pm$ 0.22 \%  & 77.74\\
171 & MOS-H & 157& 10.43 &10.08 $\pm$ 0.59 & 10.04 $\pm$ 0.59 \%  & 54.50\\
\tableline
259 & MOS-S & 128& 28.85 & 1.17 $\pm$ 0.08 &  3.42 $\pm$ 0.24 \%  & 78.01\\
259 & MOS-M & 128& 18.57 & 4.30 $\pm$ 0.28 &  6.55 $\pm$ 0.43 \%  & 68.95\\ 
259 & MOS-H & 128&  7.69 &15.73 $\pm$ 1.01 & 12.54 $\pm$ 0.81 \%  & 45.61\\
\enddata
\end{deluxetable}

\begin{deluxetable}{lccccccc}
\tablewidth{5.2in}
\tablenum{3}
\tablecaption{Cross-Correlation Results \label{tab3}}
\small
\tablehead{
\colhead{} & \colhead{} & \colhead{Band 1} & \colhead{Band 2} &
\colhead{DCF} & \colhead{DCF} & \colhead{ICF} & \colhead{ICF} \\
\colhead{Rev.} & \colhead{Instr.} & \colhead{(keV)} & \colhead{(keV)} &
\colhead{$r_{max}$} & \colhead{$\tau$ (hr)} & \colhead{$r_{max}$} & 
\colhead{$\tau$ (hr)} }
\startdata
84  & pn & 0.1-0.75 & 0.9-1.7 & 0.99 & 0.00   & 0.99 & +0.04  $\pm$ 0.05 \\
84  & pn & 0.1-0.75 & 2-10    & 0.98 & --0.08 & 0.99 & --0.04 $\pm$  0.04\\
84  & pn & 0.9-1.7 & 2-10     & 0.98 & --0.08 & 0.99 & --0.04 $\pm$  0.04\\
\tableline
165 & pn & 0.1-0.75 & 0.9-1.7 & 0.96 & --0.08 & 0.96 & --0.04 $\pm$  0.08\\
165 & pn & 0.1-0.75 & 2-10    & 0.83 & --0.75 & 0.83 & --0.71 $\pm$  0.52\\
165 & pn & 0.9-1.7 & 2-10     & 0.88 & --0.17 & 0.89 & --0.21 $\pm$  0.18\\
\tableline
171 & pn & 0.1-0.75 & 0.9-1.7 & 0.24 & --6.00 & 0.29 & +6.42  $\pm$ 5.75\\
171 & pn & 0.1-0.75 & 2-10    & 0.12 & +6.33  & 0.17 & +6.42  $\pm$ 5.65\\
171 & pn & 0.9-1.7 & 2-10     & 0.87 & 0.00   & 0.89 & +0.04  $\pm$ 0.08\\
\tableline
259 & pn & 0.1-0.75& 0.9-1.7  & 0.90 & +0.08  &  0.91 &  +0.13 $\pm$ 0.06\\
259 & pn & 0.1-0.75& 2-10     & 0.84 & +0.17  &  0.85 &  +0.13 $\pm$ 0.07\\
259 & pn & 0.9-1.7 & 2-10     & 0.96 & 0.00   &  0.96 &  +0.04 $\pm$ 0.04\\
\enddata
\tablecomments{A positive lag indicates the softer band leading
the harder band (hard band is delayed with respect to soft band).
DCF bin size is the resolution of the lightcurve,
300 sec.; ICF resolution is one half the DCF bin size.}
\end{deluxetable}

\begin{deluxetable}{lcccccccc}
\tablewidth{0pt}
\tablenum{4}
\tablecolumns{9}
\tablecaption{ Spectral Fits \label{tab4}}
\tabletypesize{\scriptsize}
\tablehead{
\colhead{ } & \multicolumn{4}{c}{EPIC-pn} &
\multicolumn{4}{c}{EPIC-MOS} \\
\colhead{ Bin } &
\colhead{ $\Gamma$ } &
\colhead{ Norm } &
\colhead{ $\rm Flux^{b}$ } &
\colhead{ $\chi^{2}_{\nu}$ } &
\colhead{ $\Gamma$ } &
\colhead{ Norm } &
\colhead{ $\rm Flux^{b}$ } &
\colhead{ $\chi^{2}_{\nu}$ } \\
\colhead{} &
\colhead{} &
\colhead{ (1 keV) } &
\colhead{ 2.0-10.0 } &
\colhead{} &
\colhead{} &
\colhead{ (1 keV) } &
\colhead{ 2.0-10.0 } &
\colhead{} }
\startdata
\cutinhead{Orbit 84}
1 & $2.24^{+0.02}_{-0.02}$ & $0.185^{+0.004}_{-0.004}$ & 
3.34 & 1.05 & 
- & - & 
- & - \\
2 & $2.22^{+0.02}_{-0.02}$ & $0.156^{+0.004}_{-0.004}$ & 
2.92 & 0.97 & 
- & - & 
- & - \\
3 & $2.22^{+0.02}_{-0.02}$ & $0.132^{+0.003}_{-0.003}$ & 
2.45 & 1.04 & 
- & - & 
- & - \\
4 & $2.20^{+0.02}_{-0.02}$ & $0.120^{+0.003}_{-0.003}$ & 
2.29 & 1.16 & 
- & - & 
- & - \\
5 & $2.20^{+0.02}_{-0.02}$ & $0.124^{+0.003}_{-0.003}$ & 
2.39 & 1.02 & 
- & - & 
- & - \\
6 & $2.27^{+0.02}_{-0.02}$ & $0.134^{+0.003}_{-0.003}$ & 
2.34 & 1.02 & 
- & - & 
- & - \\
\cutinhead{Orbit 165}
1 & $2.55^{+0.03}_{-0.03}$ & $0.063^{+0.002}_{-0.002}$ & 
0.73 & 0.99 & 
$2.53^{+0.04}_{-0.04}$ & $0.061^{+0.003}_{-0.003}$ & 
0.73 & 0.96\\
2 & $2.55^{+0.03}_{-0.03}$ & $0.057^{+0.002}_{-0.002}$ & 
0.66 & 1.01 & 
$2.51^{+0.05}_{-0.05}$ & $0.055^{+0.003}_{-0.003}$ & 
0.67 & 1.09\\
3 & $2.50^{+0.03}_{-0.03}$ & $0.055^{+0.002}_{-0.002}$ & 
0.69 & 1.10 & 
$2.46^{+0.04}_{-0.04}$ & $0.053^{+0.002}_{-0.002}$ & 
0.71 & 1.08\\
4 & $2.52^{+0.03}_{-0.03}$ & $0.060^{+0.002}_{-0.002}$ & 
0.74 & 1.02 & 
$2.47^{+0.04}_{-0.04}$ & $0.058^{+0.003}_{-0.003}$ & 
0.77 & 0.98\\
5 & $2.50^{+0.03}_{-0.03}$ & $0.067^{+0.002}_{-0.002}$ & 
0.83 & 0.94 & 
$2.42^{+0.03}_{-0.03}$ & $0.062^{+0.002}_{-0.003}$ & 
0.87 & 0.87\\
6 & $2.47^{+0.03}_{-0.03}$ & $0.067^{+0.002}_{-0.002}$ & 
0.87 & 1.05 & 
$2.46^{+0.03}_{-0.03}$ & $0.066^{+0.003}_{-0.003}$ & 
0.87 & 0.98\\
7 & $2.51^{+0.03}_{-0.03}$ & $0.066^{+0.002}_{-0.002}$ & 
0.81 & 1.03 & 
$2.53^{+0.03}_{-0.04}$ & $0.070^{+0.003}_{-0.003}$ & 
0.83 & 1.09\\
\cutinhead{Orbit 171}
1 & $2.20^{+0.02}_{-0.02}$ & $0.222^{+0.005}_{-0.005}$ & 
4.23 & 1.16 & 
$2.19^{+0.02}_{-0.02}$ & $0.213^{+0.005}_{-0.005}$ & 
4.14 & 1.02\\
2 & $2.20^{+0.02}_{-0.02}$ & $0.218^{+0.004}_{-0.005}$ & 
4.18 & 1.03 & 
$2.20^{+0.02}_{-0.02}$ & $0.216^{+0.005}_{-0.005}$ & 
4.16 & 1.16\\
3 & $2.23^{+0.02}_{-0.02}$ & $0.211^{+0.004}_{-0.005}$ & 
3.89 & 1.07 & 
$2.22^{+0.02}_{-0.02}$ & $0.206^{+0.005}_{-0.005}$ & 
3.82 & 1.23\\
4 & $2.37^{+0.02}_{-0.02}$ & $0.229^{+0.005}_{-0.005}$ & 
3.47 & 1.05 & 
$2.33^{+0.02}_{-0.02}$ & $0.221^{+0.005}_{-0.005}$ & 
3.51 & 1.21\\
5 & $2.39^{+0.02}_{-0.02}$ & $0.224^{+0.005}_{-0.005}$ & 
3.30 & 1.06 & 
$2.36^{+0.02}_{-0.02}$ & $0.220^{+0.005}_{-0.005}$ & 
3.37 & 0.94\\
6 & $2.39^{+0.02}_{-0.02}$ & $0.224^{+0.005}_{-0.005}$ & 
3.29 & 1.00 & 
$2.38^{+0.02}_{-0.02}$ & $0.225^{+0.005}_{-0.006}$ & 
3.35 & 0.90\\
7 & $2.42^{+0.02}_{-0.02}$ & $0.223^{+0.005}_{-0.005}$ & 
3.13 & 0.97 & 
$2.40^{+0.02}_{-0.02}$ & $0.219^{+0.005}_{-0.006}$ & 
3.17 & 1.01\\
8 & $2.47^{+0.02}_{-0.02}$ & $0.213^{+0.005}_{-0.005}$ & 
2.78 & 1.09 & 
$2.45^{+0.02}_{-0.02}$ & $0.212^{+0.006}_{-0.006}$ & 
2.83 & 1.15\\
9 & $2.49^{+0.02}_{-0.02}$ & $0.223^{+0.005}_{-0.005}$ & 
2.85 & 1.11 & 
$2.44^{+0.02}_{-0.02}$ & $0.221^{+0.006}_{-0.006}$ & 
3.01 & 1.12\\
\cutinhead{Orbit 259}
1 & $2.37^{+0.02}_{-0.02}$ & $0.145^{+0.003}_{-0.004}$ & 
2.16 & 1.05 & 
$2.38^{+0.03}_{-0.03}$ & $0.148^{+0.004}_{-0.004}$ & 
2.20 & 0.95\\
2 & $2.25^{+0.02}_{-0.02}$ & $0.147^{+0.003}_{-0.003}$ & 
2.64 & 1.11 & 
$2.26^{+0.02}_{-0.02}$ & $0.156^{+0.004}_{-0.004}$ & 
2.74 & 1.10\\
3 & $2.21^{+0.02}_{-0.02}$ & $0.148^{+0.003}_{-0.003}$ & 
2.79 & 1.02 & 
$2.24^{+0.02}_{-0.02}$ & $0.155^{+0.004}_{-0.004}$ & 
2.81 & 1.09\\
4 & $2.20^{+0.02}_{-0.02}$ & $0.141^{+0.003}_{-0.003}$ & 
2.70 & 1.03 & 
$2.24^{+0.02}_{-0.02}$ & $0.150^{+0.004}_{-0.004}$ & 
2.73 & 1.05\\
5 & $2.20^{+0.02}_{-0.02}$ & $0.148^{+0.003}_{-0.003}$ & 
2.87 & 0.93 & 
$2.19^{+0.02}_{-0.02}$ & $0.154^{+0.004}_{-0.004}$ & 
2.97 & 1.16\\
6 & $2.23^{+0.02}_{-0.02}$ & $0.157^{+0.003}_{-0.003}$ & 
2.86 & 1.00 & 
$2.24^{+0.02}_{-0.02}$ & $0.160^{+0.004}_{-0.004}$ & 
2.90 & 1.00\\
7 & $2.34^{+0.02}_{-0.02}$ & $0.145^{+0.004}_{-0.004}$ & 
2.27 & 1.13 & 
$2.38^{+0.03}_{-0.03}$ & $0.156^{+0.005}_{-0.005}$ & 
2.31 & 1.02\\
\enddata
\tablecomments{a) All errors are $1 \sigma$ for two interesting 
parameters ($\delta \chi^{2} = 2.30$). b) Deabsorbed flux in units of 
$\rm 10^{-10} \ erg \ cm^{-2} \ s^{-1}$}. 
\end{deluxetable}

\newpage

\begin{figure}
\epsfig{file=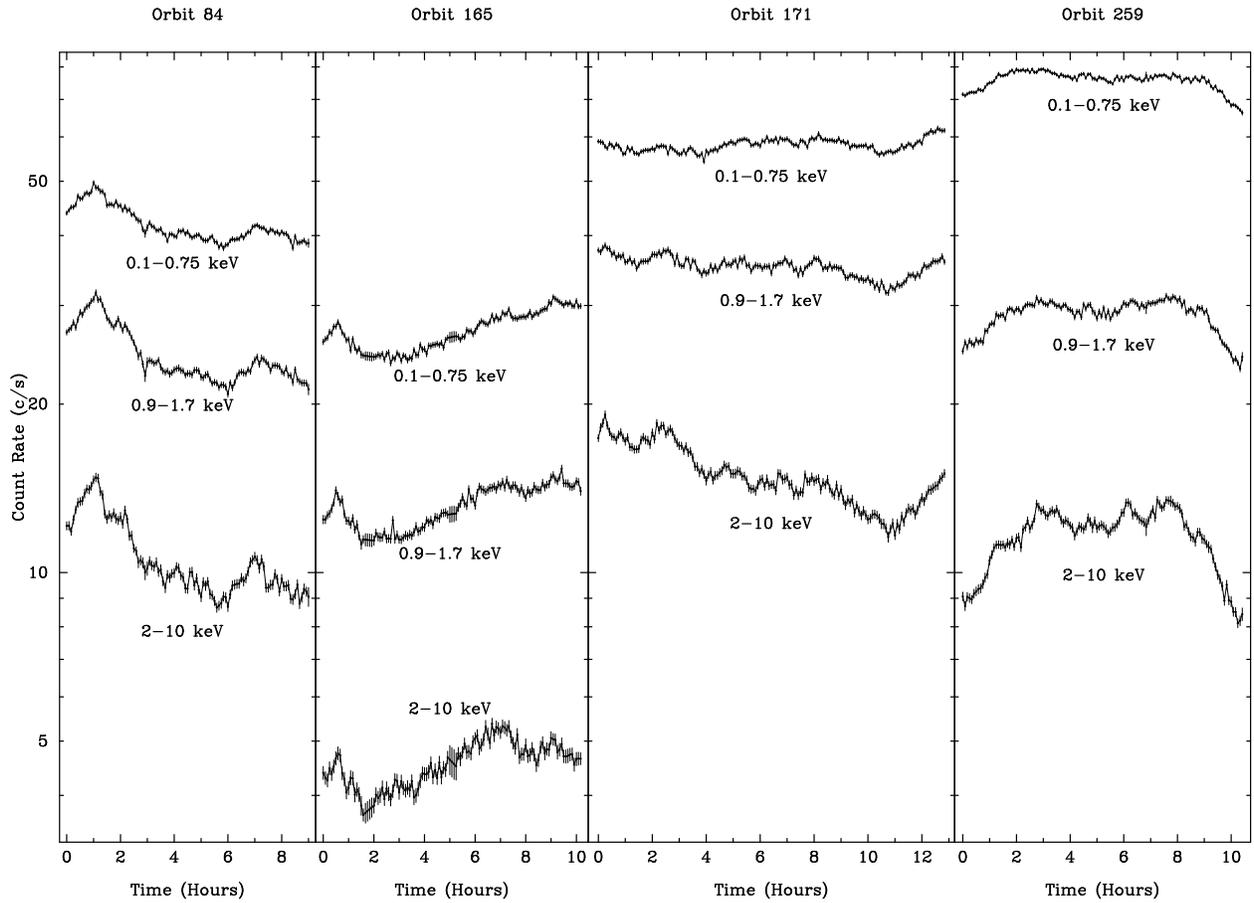,angle=270,width=\columnwidth}
\caption{Grid of observed \xmm\ pn light curves for \mkn.
The Orbit 84 data are shown on the left (2000 May 25) through to Orbit 
259 (2001 May 8) on the right.}
\end{figure}

\newpage

\begin{figure}
\epsfig{file=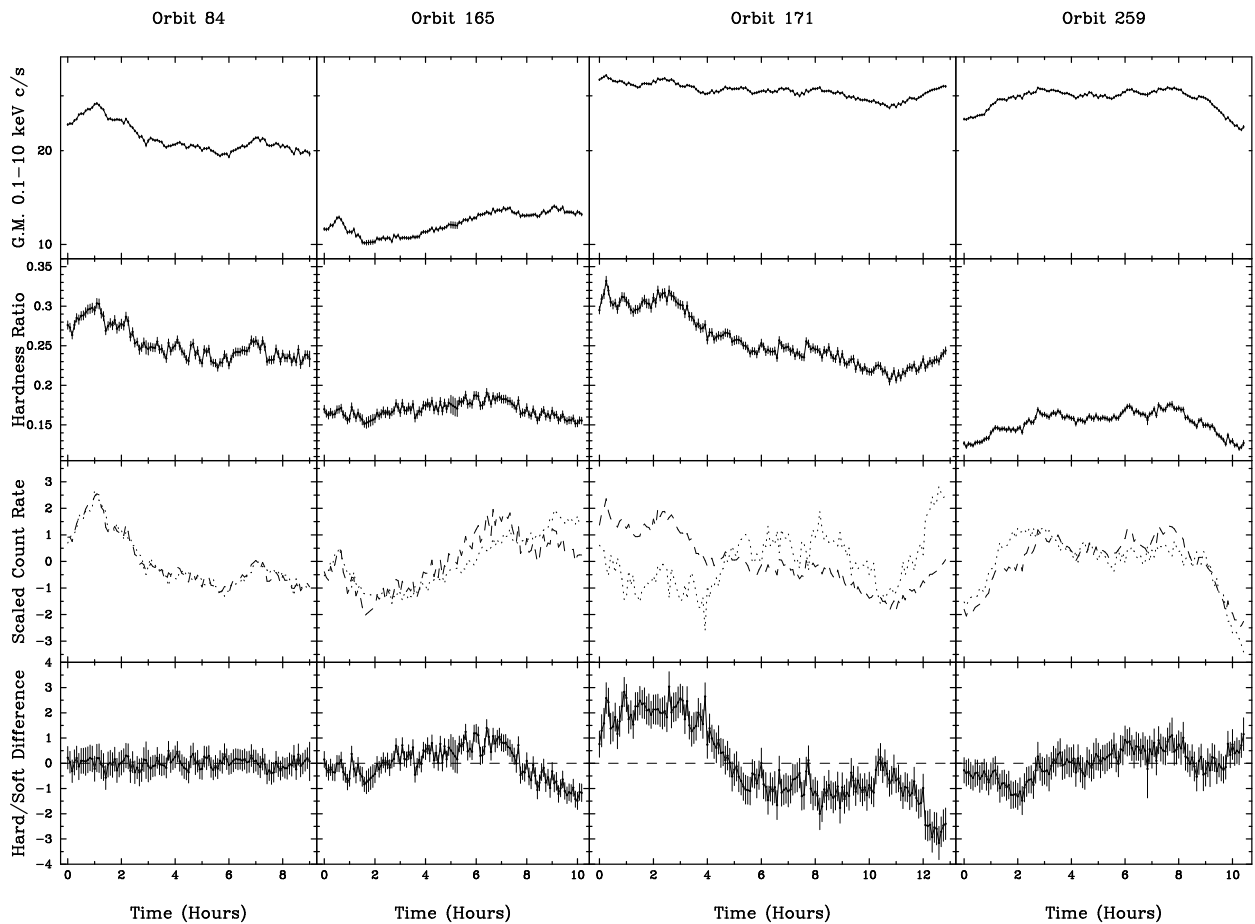,angle=270,width=\columnwidth}
\caption{Grid of light curves of derived quantities.
From the top, the light curves shown are the geometric mean count rate,
2--10~keV/0.1--0.75~keV hardness ratio, and overplot diagrams and
difference diagrams (see text for details). In the overplot diagrams the 
soft band is denoted by the dotted line and tha hard band is denoted by 
the dashed line. Note in particular that the hard and soft light curves 
track very well during O84 and O259, indicating a good correlation between 
bands, but not during O165 and O171, indicating that those data are less
well-correlated.}
\end{figure}

\newpage

\begin{figure}
\epsfig{file=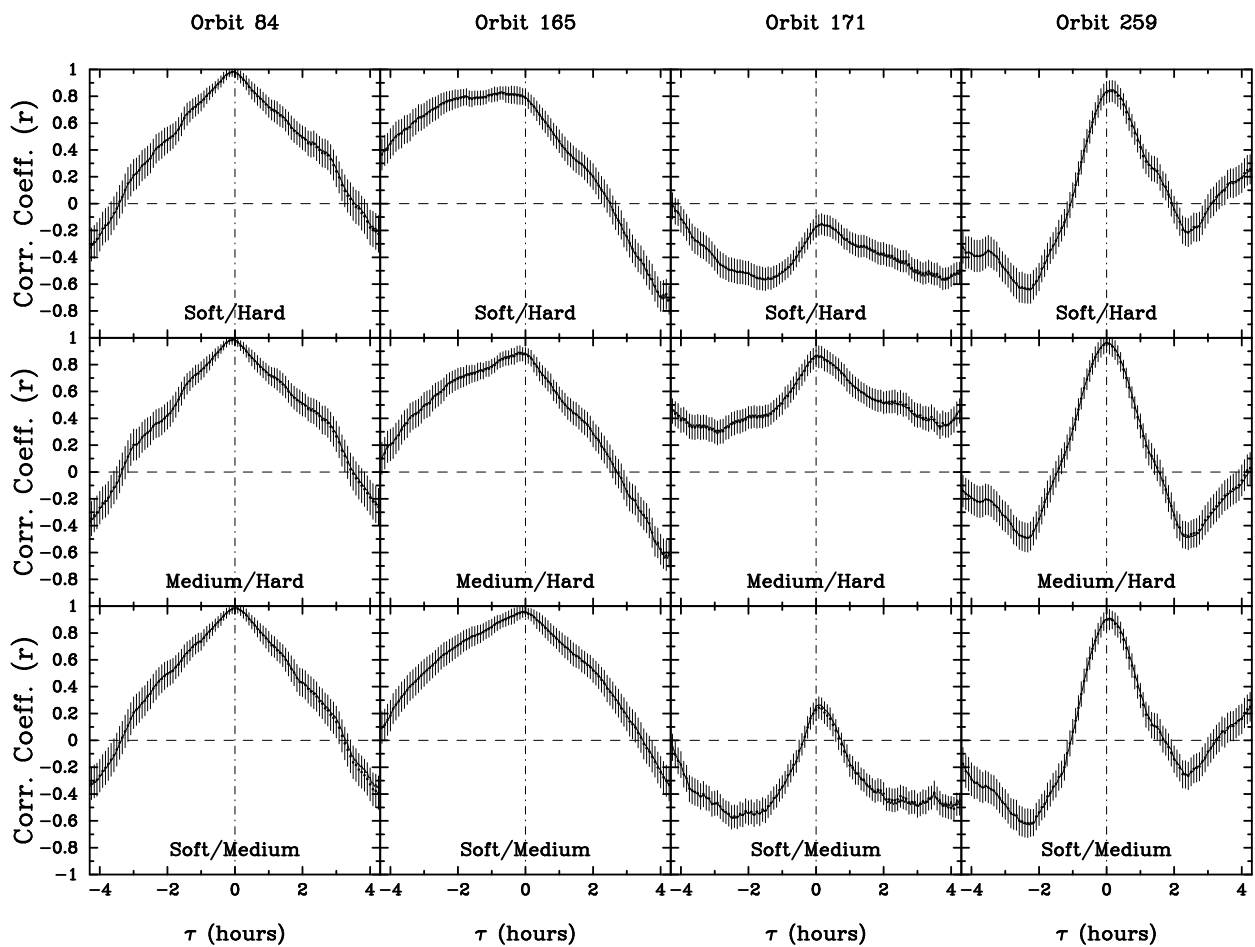,angle=270,width=\columnwidth}
\caption{Grid of interband CCFs. The hard/soft band CCFs are shown on the 
top, hard/medium band CCFs in the middle, and medium/soft band CCFs on the 
bottom. Only the O84 and O259 data show a usable correlation, and those data 
are consistent with zero lag (to 2$\sigma$ limits of $\ls$0.3~ksec). }
\end{figure}

\newpage

\begin{figure}
\epsscale{0.8}
\plotone{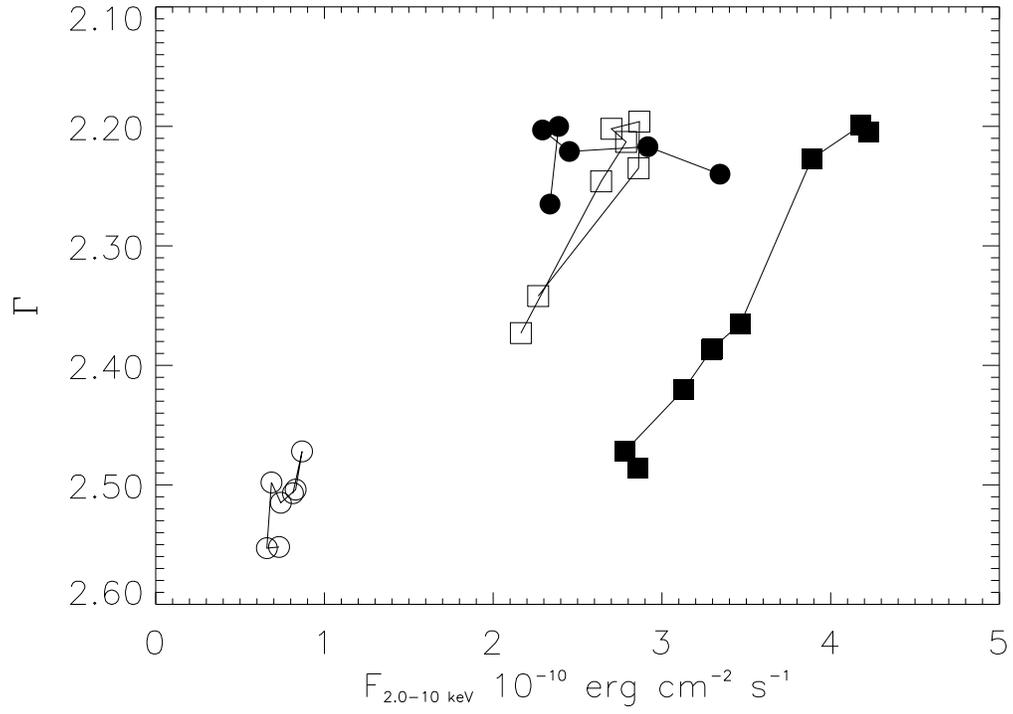}
\caption{EPIC-pn photon spectral index in the 2.0-10 keV band plotted against 
the deabsorbed 2.0-10 keV flux. O84 are the closed circles. O165 are the open 
circles. O171 are the closed squares and O259 are the open squares.}
\end{figure}

\end{document}